\documentclass[a4paper,fleqn]{cas-dc}
\usepackage{algorithm,algorithmic}
\usepackage{units}
\usepackage{float}

\usepackage[round,authoryear]{natbib}

\def\tsc#1{\csdef{#1}{\textsc{\lowercase{#1}}\xspace}}
\tsc{WGM}
\tsc{QE}

\begin{document}
\let\WriteBookmarks\relax
\def\floatpagepagefraction{1}
\def\textpagefraction{.001}

\shorttitle{Decision-based iterative fragile watermarking for model integrity verification}    

\shortauthors{Zhaoxia Yin, Heng Yin, Hang Su, Xinpeng Zhang, Zhenzhe Gao}  

\title [mode = title]{Decision-based iterative fragile watermarking for model integrity verification}  

\author[1]{Zhaoxia Yin}[type=editor,
style=chinese,
orcid=0000-0003-0387-4806]
\ead{zxyin@cee.ecnu.edu.cn}
\address[1]{Shanghai Key Laboratory of Multidimensional Information Processing,\
	East China Normal University\
	Shanghai,\
	200241, \
	Shanghai,\
	China}
\author[2]{Heng Yin}[type=editor,
style=chinese,
orcid=0000-0001-8900-649X]
\ead{e21301335@stu.ahu.edu.cn}
\address[2]{Anhui Provincial Key Laboratory of Multimodal Cognitive Computation, Anhui University,\
	Hefei,\
	230000, \
	Anhui,\
	China}
\author[3]{Hang Su}[type=editor,
style=chinese,
orcid=0000-0001-8294-6315]
\fnmark[*]
\ead{suhangss@mail.tsinghua.edu.cn}
\cortext[1]{Corresponding author}
\address[3]{Department of Computer Science and Technology, Tsinghua University,\
	Beijing,\
	100000,\ 
	Beijing,\
	China}
\author[4]{Xinpeng Zhang}[type=editor,
style=chinese,
orcid=0000-0001-5867-1315]
\ead{zhangxinpeng@fudan.edu.cn}
\address[4]{School of Computer Science and Technology, Fudan University,\
	Shanghai,\
	200241,\ 
	Shanghai,\
	China}
\author[1]{Zhenzhe Gao}[type=editor,
style=chinese,
orcid=0009-0008-2545-796X]
\ead{51255904049@stu.ecnu.edu.cn}
\begin{abstract}
	With the development of artificial intelligence, fine-tuning pretrained models has become increasingly accessible for average users, enabling them to easily adapt models to their specific tasks. 
	As a result, the practical applications of such models have been further accelerated across various fields. 
	Typically, foundation models are hosted on cloud servers to meet the high demand for their services. However, this exposes them to security risks, as attackers can modify them after uploading to the cloud or transferring from a local system. 
	To address this issue, we propose an iterative decision-based fragile watermarking algorithm that transforms normal training samples into fragile samples that are sensitive to model changes. 
	We then compare the output of sensitive samples from the original model to that of the compromised model during validation to assess the model's completeness.
	The proposed fragile watermarking algorithm is an optimization problem that aims to minimize the variance of the predicted probability distribution outputed by the target model when fed with the converted sample.
	We convert normal samples to fragile samples through multiple iterations. 
	Our method has some advantages: (1) the iterative update of samples is done in a decision-based black-box manner, relying solely on the predicted probability distribution of the target model, which reduces the risk of exposure to adversarial attacks, (2) the small-amplitude multiple iterations approach allows the fragile samples to perform well visually, with a PSNR of 55 dB in TinyImageNet compared to the original samples, (3) even with changes in the overall parameters of the model of magnitude 1e-4, the fragile samples can detect such changes, and (4) the method is independent of the specific model structure and dataset. We demonstrate the effectiveness of our method on multiple models and datasets, and show that it outperforms the current state-of-the-art.
\end{abstract}


\maketitle

\section{Introduction}\label{intro}
In the current era of information explosion, the remarkable improvement in computing power has led to the increasing popularity of machine learning algorithms. This can be attributed to the significant progress made in the field of machine learning algorithms, coupled with the availability of convenient tools provided by major companies such as TensorFlow, MxNet, and PyTorch. As a result of these advancements, the performance of deep neural networks (DNNs) in image recognition \citep{imgrec1,imgrec2,imgrec3}, natural language processing \citep{nlp1,nlp2,nlp3}, and speech recognition \citep{speech1,speech2} has reached the state-of-the-art (SOTA) level.
\begin{figure}[htbp]
	\centering
	\includegraphics[width=0.5\textwidth]{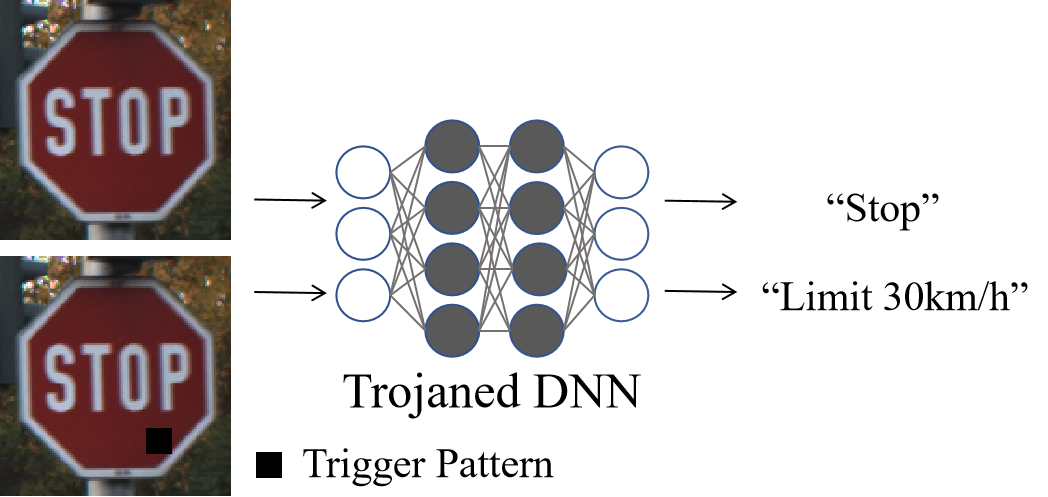}
	\caption{A model injected with a backdoor will output a specific result as intended by the attacker, once it recognizes a specified trigger pattern.}
	\label{FIG:1}
\end{figure}
\par
In the meantime, the field of Artificial Intelligence (AI) is currently undergoing a paradigm shift with the emergence of foundation models \citep{Bigmodel}. These models are large pretrained models that can be easily adapted to perform well on a wide variety of tasks through fine-tuning or prompting. However, the training of these foundation models requires enormous amounts of data, powerful computing hardware resources, significant amounts of time, and experienced experts, which can be challenging for small-sized enterprises.
As a result, many ordinary enterprises and researchers have opted to transfer foundation models to their applications and research through low-cost technologies such as fine-tuning.
\par
\begin{figure*}[htbp]
	\centering
	\includegraphics[width=\textwidth]{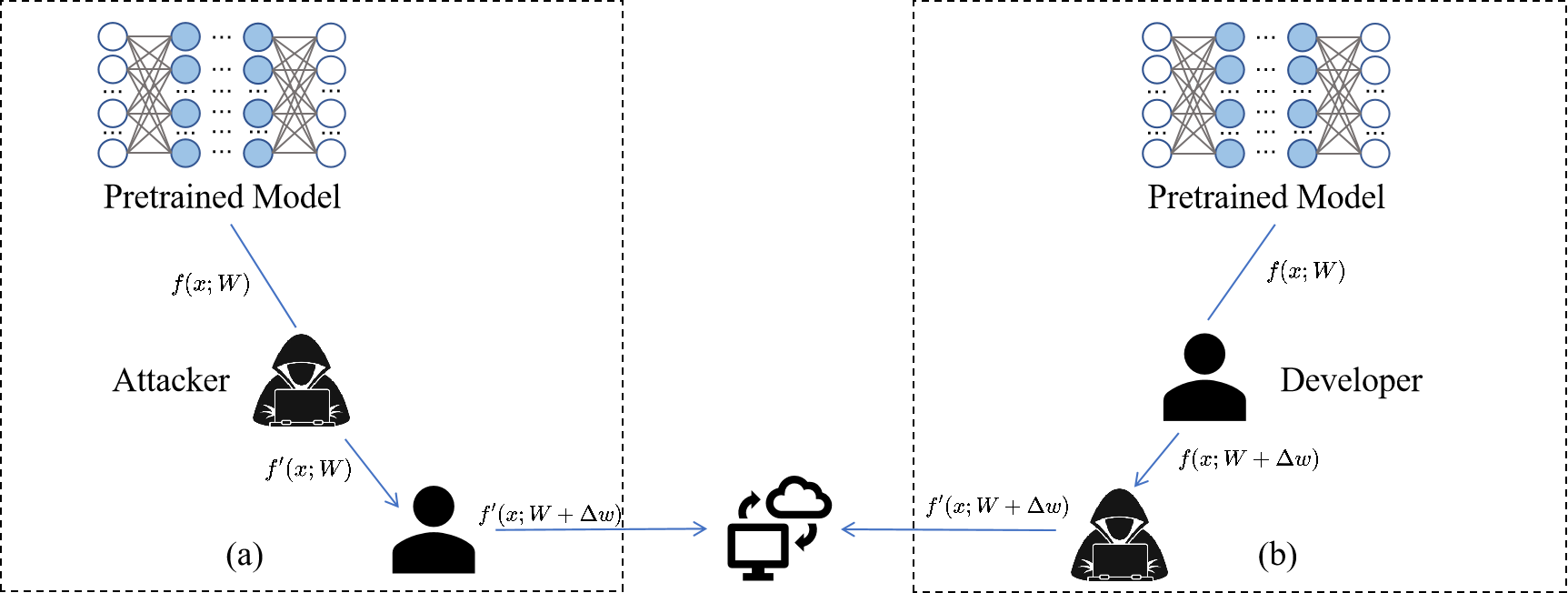}
	\caption{This figure illustrates two scenarios in which an attacker can compromise the integrity of a model by injecting a backdoor. In Figure (a), the attacker can inject a backdoor into the pretrained model $f(\mathbf{x} ;W)$ and fine-tune it to obtain the backdoored model $f'(\mathbf{x} ;W)$, which is then transferred to the user when the pretrained model $f(\mathbf{x} ;W)$ is distributed. The model with the backdoor appears to function normally in the vast majority of inputs, making it difficult for the user to detect any anomalies. The user then fine-tune $f'(\mathbf{x} ;W)$ for their specific task and deploy it on a remote host or cloud server. In Figure (b), the attacker can inject a backdoor into the model when the user deploys it on a remote host or cloud-platform.}
	\label{FIG:2}
\end{figure*}
While pretrained models offer convenience, they also face certain security risks. Figure \ref{FIG:1} illustrates a scenario where the model is used with a backdoor present. 
In this example, the adversary deliberately adds a trigger pattern into the traffic sign "Stop" to make the model recognize it as "Speed Limit".
\par
Typically, the foundation model needs to be hosted on a cloud platform due to its large-scale, and the Application Programming Interface (API) is released to provide services. Figure \ref{FIG:2} shows two scenarios where the attacker injects a backdoor. One scenario is where users obtain pretrained models that have been tampered with by attackers in the process of transmission, while the other is where attackers directly tamper with models deployed on remote machines.
A common tampering technique is the DNN backdoor attack, also known as a Trojan \citep{trojan1,trojan2,trojan3,datapoison1,datapoison2,datapoison3,datapoison4,hardware1,hardware2,hardware3}. The backdoor attack modifies the model parameters by fine-tuning to inject hidden patterns or calculates the adversarial weight perturbation through the gradient information of the model, then attaches it directly to the original parameters. The tampered classifier can predict well in most inputs but will behave abnormally in some specified triggers. 
When models with backdoor patterns are deployed in security-critical applications, such as autopilot, they can cause significant safety accidents.
\par 
It is essential to have a mechanism to verify the integrity of the model both before transferring the foundation model to our own task and when deploying the model on a cloud platform. However, since the model hosted on the cloud platform is also a black-box for the developer, verifying the model's integrity hosted on the cloud platform is a challenging task. Therefore, we can only verify the model's integrity by querying it once it is deployed on the cloud platform. In this black-box verification setting, the most plausible way to verify the integrity of the hosted model is to check if the prediction results output by the original model and the hosted model match. A mismatch in the prediction results indicates a compromise.
Extensively querying a model in the cloud using each instance of the training dataset is very expensive since models are typically hosted on a pay-per-query basis. Moreover, adversaries often make slight modifications to a hosted model to avoid detection, making the prediction performance of the compromised model nearly identical to the original model. Therefore, detecting integrity by querying the prediction results of each training sample when the dataset is large is not practical.
Since small changes to the model do not affect the prediction results for most training samples, we can look for samples that are highly sensitive to changes in the model. If the model undergoes malicious fine-tuning or similar alterations, the prediction results for these sensitive samples will change.
\par
In this paper, we propose a decision-based \citep{score-black} black-box fragile watermarking algorithm that makes normal training samples sensitive to model modifications. Our approach is inspired by the observation in \citep{hardsample} that there are hard-to-train samples in the training set that are repeatedly remembered and forgotten during model training. We refer to these samples as fragile samples and analyze their characteristics. These fragile samples have much smaller variances in the prediction probability distribution of the target model compared to the easy-to-learn samples with distinct features. Therefore, we use the variance of the predicted probability distribution of the normal samples in the target model as a loss to iteratively update the samples to make them sensitive to model modification. The goal of the watermarking method is achieved by making these normal training samples fragile and composing the set of triggers.
The process of making the samples fragile only requires the predicted probability distribution provided by the target model, which greatly reduces the risk of exposing the model parameters and triggering adversarial attacks during the watermarking process. 
Additionally, small iterations of samples in each round result in fragile samples with a Peak signal-to-noise ratio (PSNR) of 55 dB obtained in TinyImageNet, reducing the risk of fragile samples being discovered by attackers.
Furthermore, experiments show that our method is simple in process and efficient in detection, even for changes of only 1e-4 magnitude in the overall parameters.
The key contributions of this paper are summarized as follows:
\begin{itemize}
	\item[$\bullet$] We propose a novel decision-based black-box fragile watermarking algorithm. 
	This algorithm only requires the predicted probability distribution of the final output of the target model to start optimizing the sample, further reducing the risk of adversarial attacks due to the exposure of model parameter information.
	\item[$\bullet$] This algorithm optimizes the normal samples by minimizing the variance of their predicted probability in the target model, and the final PSNR of the obtained fragile samples in TinyImageNet can reach 55dB.
	\item[$\bullet$] Sufficient experiments have demonstrated the effectiveness of our method, which can detect even modifications of only 1e-4 magnitude. And the method is independent of specific models and datasets.
\end{itemize} 
\section{Related Work and Background}\label{related}
In this paper, we focus on model watermarking for verifying the integrity of image classification models.
\subsection{Model Watermarking}
The technology of model watermarking is derived from traditional digital watermarking \citep{digitalwatermark,digit2,digit3,digit4} and was first proposed by \citep{traditional1}. 
Over the years, model watermarking technology has been developed and can be classified into two main categories: inserting the watermark or related information into the model parameters or constructing a trigger that can make the watermarking model output specific predictions. Initially, researchers designed robust model watermarking methods \citep{traditional1,traditional2,traditional3,traditional4,traditional5,traditional6,traditional7,traditional8,traditional9,traditional10} to protect the copyright information of the model and enable illegal users to identify stolen models. Later, researchers explored fragile model watermarking \citep{fragile1,fragile2,fragile3} technology to verify the integrity of model. 
Fragile model watermarking is sensitive to model fine-tuning and tiny malicious model parameter modifications. 
Meanwhile, model fingerprint \citep{fingerprint1,fingerprint2,fingerprint3,fingerprint4,fingerprint5,fingerprint6} technology has also been proposed. The core of fingerprint technology is to make fragile samples as triggers.
Therefore, it can still be classified as a category of fragile model watermarking technology.
\par
\citep{fragile1,fragile2} proposed a white-box model integrity verification method by embedding watermarks into model parameters. 
Although these methods can effectively detect modifications to the watermark embedding layer, white-box model integrity verification techniques fail in a black-box verification setting where internal parameter information of models deployed on cloud platforms cannot be accessed. Therefore, we can only verify the integrity of the model by querying it. The following techniques all verify the integrity of the model in a black-box manner.
\citep{fragile3} proposed a fragile watermark method that embeds watermarks by fine-tuning the target model. Because the model needs to be fine-tuned to embed the watermark pattern, it still has some potential impact on the performance of the watermark model.
\citep{fingerprint1,fingerprint2,fingerprint3,fingerprint4,fingerprint5} proposed using a carefully crafted set of samples and corresponding labels as a trigger set to watermark the model. These methods detect model modifications by observing the differences between predicted labels in the original model and the modified model. The process of creating fragile samples requires detailed gradient information of the model parameters. But if the repeated watermarking process leads to the exposure of parameter information, attackers can carefully create adversarial examples based on the parameter information, leading to another type of risk.
\par
Our algorithm only requires the predicted probability distribution of the final output of the target model to start optimizing the sample.
Previous methods of the same type generate noise-like images, which are easily detected and filtered out. 
Proposed method updates directly on the original training samples to get fragile sample.
The fragile sample has small visual differences with the original images and experimentally proven to be more sensitive to model modifications.
\subsection{Deep Neural Networks}
Deep Neural Networks are functions of the form $y = f(x;W)$ that can map an input $x \in \mathcal{X}$ to an output $y \in \mathcal{Y}$, where the function is parameterized by $W$.
The DNN model typically consists of multiple computation layers, and the final layer is usually a fully-connected layer that outputs the predicted scores for each category.
The primary objective of training a DNN model is to discover the optimal parameters $W$ that accurately capture the mapping relationship between $\mathcal{X}$ and $\mathcal{Y}$. Typically, the training dataset $D^{train}=\{x_{i}^{train},y_{i}^{train}\}_{i=1}^{N}$ comprises N images, where $x_{i}^{train} \in \mathcal{X}$ denotes the $i$-$th$ input image and $y_{i}^{train}$ corresponds to its ground-truth label. During the training phase, the objective is to minimize the loss function $L$ with respect to $W$:
\begin{center}
	\begin{equation}\label{eq1}
		W^{*}=arg\mathop{min}\limits_{W}(\sum_{i=1}^{N}L(y_{i}^{train},f(x_{i}^{train};W))),
	\end{equation}
\end{center}
\begin{figure*}[htbp]
	\centering
	\includegraphics[width=\textwidth]{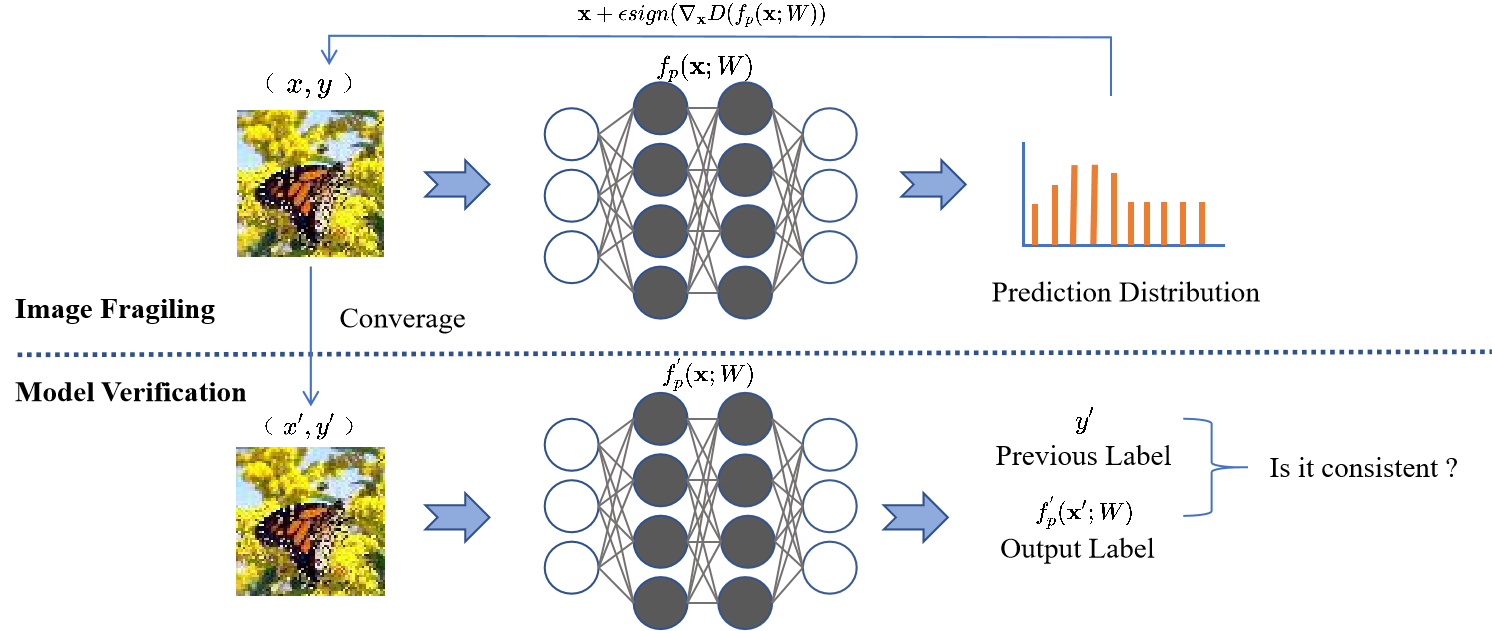}
	\caption{The general framework of our approach. Initially, we select a random sample $(\bold{x},y)$ from the training dataset and input it into the target model $f_{p}(\mathbf{x} ;W)$ for watermarking purposes, which yields a predicted probability distribution. 
		We then calculate the variance of the predicted probability distribution, and use this variance value as a loss to iteratively convert the original sample into a fragile sample $(\bold{x}',y')$ until convergence is achieved.
		To verify whether model $f_{p}(\mathbf{x} ;W)$ is consistent with model $f'_{p}(\mathbf{x'} ;W)$, we compare the predicted result of the original sample, denoted as $f'_{p}(\mathbf{x'} ;W)$, to that of the fragile sample, denoted as $y'$. 
		If  $y'$ is consistent with  $f'_{p}(\mathbf{x'} ;W)$, then the model's integrity has not been compromised. 
		If, on the other hand,  $y'$ and  $f'_{p}(\mathbf{x'} ;W)$ differ significantly, then the model has been altered.
		This process allows us to actively optimize normal samples to become fragile samples, and then use these fragile samples to detect any unauthorized changes to the model.}
	\label{FIG:3}
\end{figure*}
This optimization process measures the errors between the predicted and actual labels and updates the parameters $W$ accordingly.
\subsection{Model Integrity Attacks}
The primary threats to the integrity of machine learning models are backdoor injection and poisoning via malicious data during fine-tuning. These attacks are significant because they can lead to incorrect predictions and ultimately pose security risks. While other factors such as compression and random noise can also impact model integrity to some extent, attackers cannot effectively leverage these factors to achieve their objectives.
\par
{\bf{Backdoor Injection.}} 
Backdoor injection is an attack strategy that aims to implant a trigger pattern into a model, causing it to misclassify samples containing a particular trigger.
To execute this attack, the attacker starts with a pretrained DNN model and identifies specific "critical" neurons that are vital in determining the model's output.
The attacker then modifies the weights along the path from the selected neurons to the final layer by retraining the model using data that includes the trigger.
\par
{\bf{Poisoning Fine-tune.}} 
The objective of Poisoning fine-tuning is to cause the model to misclassify a specific class. The attacker accomplishes this by contaminating the dataset with specifically crafted malicious samples. There are two types of such attacks to consider: the first is the error-generic poisoning attack, where the outputs of the compromised model for the target class can be anything. The second type is the error-specific poisoning attack, where the attacker modifies the model to misclassify the target class as a particular fixed class of their choosing.
\section{Method}\label{}
Figure \ref{FIG:3} depicts the general framework of our approach for making samples fragile and utilizing these samples to verify the integrity of the model. The process involves several steps.
Initially, we select a random sample $(\bold{x},y)$ from the training dataset and input it into the target model $f_{p}(\mathbf{x} ;W)$ for watermarking purposes, which yields a predicted probability distribution. 
We then calculate the variance of the predicted probability distribution, and use this variance value as a loss to iteratively convert the original sample into a fragile sample $(\bold{x}',y')$ until convergence is achieved.
To verify whether model $f_{p}(\mathbf{x} ;W)$ is consistent with model $f'_{p}(\mathbf{x'} ;W)$, we compare the predicted result of the original sample, denoted as $f'_{p}(\mathbf{x'} ;W)$, to that of the fragile sample, denoted as $y'$. 
If  $y'$ is consistent with  $f'_{p}(\mathbf{x'} ;W)$, then the model's integrity has not been compromised. 
If, on the other hand,  $y'$ and  $f'_{p}(\mathbf{x'} ;W)$ differ significantly, then the model has been altered.
This process allows us to actively optimize normal samples to become fragile samples, and then use these fragile samples to detect any unauthorized changes to the model.
The next part of this chapter details the origins of our approach and the specific implementation details.
\subsection{Overview}\label{yhsubsec6}
\par
In previous studies, it was discovered that during model training, certain samples in the training set would repeatedly be classified correctly after one round of training, only to be misclassified in the next round. Upon analysis, it was determined that these samples were less robust than other, easier-to-learn samples. 
To analyze these hard-to-learn or sensitive samples, we trained the model and recorded their classification confidence, which revealed that the prediction probability distribution for these samples was more uniform across categories than that of easy-to-learn samples.
The fragility of these samples lies in their sensitivity to adjustments made during model training, which can cause them to be forgotten. 
However, it is impractical to collect these fragile samples as a validation trigger set to verify the integrity of pre-trained models, since we rarely have access to the training process of large pre-trained models. Instead, we use the property of fragile samples, namely their more uniform prediction probability distribution in the target model, to actively optimize normal samples to become fragile samples.
Figure 4 provides further insight into why fragile samples are sensitive to adjustments such as model fine-tuning. 
Specifically, these samples are located near the model decision boundary, meaning that their prediction result can change significantly with even slight changes in the target model.
\par
\begin{figure}[ht]
	\begin{minipage}[b]{1.0\linewidth}
		\centering
		\centerline{\includegraphics[width=8.5cm]{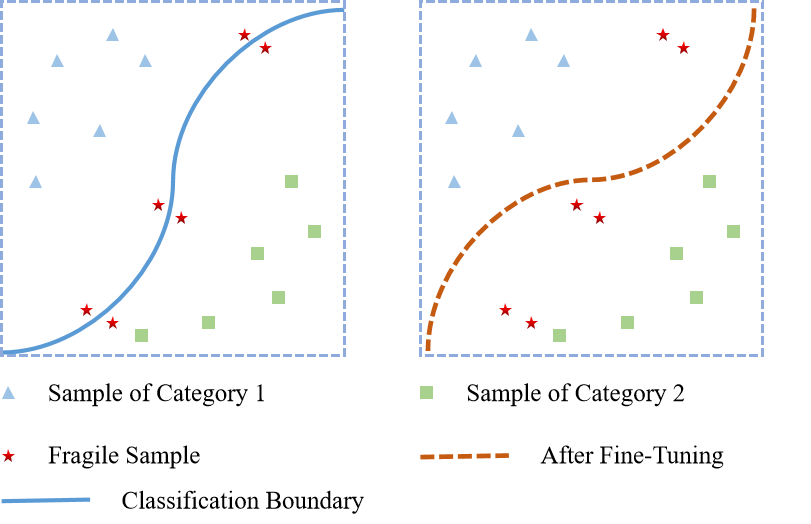}}
	\end{minipage}
	\vspace{3pt}
	\caption{Fragile samples are those located near the classification boundary of a watermarked classifier. Malicious modifications to the parameters of the watermarked classifier can lead to slight changes in the classification boundary, which in turn affects the classification results of the fragile samples. To verify the integrity of watermarked models, users can compare the difference between the predicted label and the previous label.}
	\label{FIG:4}
\end{figure}
When evaluating the generated fragile samples, we adhere to the main characteristics that were defined in the CVPR-2019 \cite{fingerprint1}:\par
{\bf{Effectiveness.}}
The fragile watermarking method should be sensitive to any malicious modifications made to the watermarked model.\par
{\bf{Efficiency.}}
When utilizing the fragile watermarking method to verify the integrity of the target model, the number of triggers should be minimized due to the cost of querying watermarked models. \par
{\bf{Black-box verification.}}
Due to the risk of exposing model parameters during white-box verification, black-box access should be used as much as possible when verifying the model's integrity.\par
{\bf{Hard to spot.}}
The fragile samples should be hard to spot and resemble natural inputs, so that adversaries cannot easily distinguish whether it is being used for integrity checking or for normal model serving.\par
{\bf{Generalizable.}}
The watermarking algorithm should be generalization and independent of machine learning models, training datasets, and attacks. It must be capable of detecting any unknown attacks.\par
These characteristics serve as the requirements for the evaluation process.
By following these requirements, we can ensure that the fragile samples are of high quality and meet the necessary standards for verifying the integrity of the model. The use of established criteria also facilitates the comparison of results across different studies and helps to promote consistency in the field.
\subsection{Fragilization Samples}
In fact the whole process of making a normal sample become fragile is very simple.
We denote the prediction probability distribution of an $n$-category task model as $\bold{p}$, where $p_i$ is the prediction probability of the $i$-th category. 
The expectation of $\bold{p}$ is first calculated using 
\begin{equation}\label{eq2}
	E(\bold{p}) = (\sum_{i=1}^{n} p_i)/{n} ,
\end{equation}
and then the variance is calculated using 
\begin{equation}\label{eq3}
	D(\bold{p}) =  \sum_{i=1}^{n}(p_i - E(\bold{p}))^2p_i .
\end{equation}
The optimization method for iterating normal samples into fragile samples is
\begin{equation}\label{eq4}
	\bold{x}' = \bold{x} + \epsilon sign(\nabla_{\bold{x}}D(f_p(\bold{x};W)),
\end{equation}
where $\bold{x}$ is the normal sample, model $f_p(\bold{x};W)$ outputs its predicted probability distribution, the direction of the gradient is calculated by the $sign()$, which is a function used to find the sign of the value, for example, for inputs greater than 0, the output is 1, for inputs less than 0, the output is -1, and for inputs equal to 0, the output is 0. We use the hyperparameter $\epsilon$ to control the magnitude of the modification at each step.\par
\begin{algorithm}[ht]
	\caption{Get fragile samples}\label{algo1}
	\begin{algorithmic}[1]
		\STATE {\bf{Function}} Get-Fragile-samples$(f_{p}(\bold{x};W),\bold{x},\epsilon,itr)$
		\STATE /* $\bold{x}:$ the normal sample from training set*/
		\STATE /* $f_p(\bold{x};W):$ the output of the model $f_p$ is the predicted probability distribution of $\bold{x}$ and its parameters are $W$*/
		\STATE /* $\epsilon:$ the hyperparameter to control the magnitude of the modification at each step*/
		\STATE /* $itr:$ maximum number of iterations*/
		\STATE $i \Leftarrow 1$
		\STATE $\bold{x'} \Leftarrow \bold{x}$
		\WHILE{$i \leqq itr$}
		\STATE $\bold{x'} = \bold{x'} + \epsilon sign(\nabla_{\bold{x'}}D(f_p(\bold{x'};W))$
		\STATE $i \Leftarrow i+1$
		\ENDWHILE
		\STATE $y' \Leftarrow Argmax(f_{p}(\bold{x'};W))$
		\STATE {\bf{return}} ($\bold{x'},y')$
	\end{algorithmic}
\end{algorithm}
Algorithm \ref{algo1} describes in more detail the process of converting normal samples to fragile samples.
In this algorithm, the inputs are the model $f_p$ to be watermarked, the normal sample $\bold{x}$, the amount of perturbation $\epsilon$ in each round and the number of desired iterations $itr$.
At each iteration, add the perturbation for that epoch in Line 9. 
After the final iteration, we get the fragile sample and the predicted label of this fragile sample in the target model.
In this way, we obtain multiple fragile samples to form a validation trigger set $(\mathcal{X'},\mathcal{Y'})$ to watermark the target model.
\subsection{Integrity Verification}\label{yh3subsec}
Next, we use the validation trigger set $(\mathcal{X'},\mathcal{Y'})$ obtained above to verify model integrity.
The validation process is described in detail in Algorithm \ref{algo2}, where we iterate through the samples in the input trigger set. 
As long as one of the samples has a prediction result that is inconsistent with the previous one, the model is judged to have been tampered with.
\begin{algorithm}[h]
	\caption{Verify the integrity of watermarked model}\label{algo2}
	\begin{algorithmic}[1]
		\STATE {\bf{Function}} Integrity-Verification$(\mathcal{X'}, \mathcal{Y'},f(\bold{x};W))$
		\STATE /*$\mathcal{X'}:$ the verification trigger set consists of fragile samples*/
		\STATE /*$\mathcal{Y'}:$ the corresponding label of $\mathcal{X'}$*/
		\STATE /*$f(\bold{x};W):$ the model watermarked with $(\mathcal{X'},\mathcal{Y'})$*/
		\STATE $\bold{x}_i \in \mathcal{X'}$
		\STATE $y_i \in \mathcal{Y'}$
		\STATE $i \Leftarrow 1$
		\WHILE{$i\leqq x.length$} 
		\IF{$f(\bold{x}_i;W)!=y_i$}
		\STATE {\bf{return}} False
		\ENDIF
		\STATE $i \Leftarrow i+1$
		\ENDWHILE
		\STATE {\bf{return}} True
	\end{algorithmic}
\end{algorithm}
\par
To assess the sensitivity of fragile samples, we adopt 
\begin{equation}\label{eq5}
	ChangeRate = \left(\sum_{i=1}^n \mathbf{1}_{condition}\left( f(\bold{x}_i;W)!=y_i\right)\right)/n,
\end{equation}
as the authentication metrics.
$\mathbf{1}_{condition}$ is a conditional function, it gets 1 when the condition is true, otherwise gets 0.
\section{Experiments}\label{experiment}
To evaluate the proposed method in this paper, we used four different datasets, MNIST \citep{mnist}, CIFAR \citep{cifar} and TinyImageNet, with small to large image sizes and categories from 10 to 200. Table \ref{Table:1} lists the specific information of these datasets. The test set size for each of these datasets is 10,000 images.
\par
\begin{table}[width=.45\textwidth,cols=4,pos=ht]
	\caption{This table shows the details of the evaluation datasets used, where MNIST is a one-dimensional set of black and white handwritten images, Cifar is a dataset containing colorful images of vehicles, animals, etc. TinyImageNet is a dataset containing colorful nature images with more detail and more categories.}
	\label{Table:1}
	\begin{tabular*}{\tblwidth}{@{} LLLLL@{} }
		\toprule
		Dataset & Size & Train & Test & classes \\
		\midrule
		MNIST & 1$\times$28$\times$28 & 60000 & 10000 & 10\\
		CIFAR10 & 3$\times$32$\times$32 & 50000 & 10000 & 10\\
		CIFAR100 & 3$\times$32$\times$32 & 50000 & 10000 & 100\\
		TinyImageNet & 3$\times$64$\times$64 & 100000 & 10000 & 200\\
		\bottomrule
	\end{tabular*}
\end{table}
For each dataset we pre-trained some models to correspond to it, and Table \ref{Table:2} shows these correspondences and the accuracy.
\begin{table}[width=.45\textwidth,cols=4,pos=ht]
	\caption{The table shows the pre-trained models corresponding to each dataset and their accuracy (acc).}
		\label{Table:2}
	\begin{tabular*}{\tblwidth}{@{} LLLL@{} }
		\toprule
		Model & Dataset & Top-1 Acc & Top-5 Acc\\
		\midrule
		Resnet18 & MNIST & 99.7 & 100.0\\
		Resnet50 & CIFAR10 & 91.3 & 99.0\\
		Resnet50 & CIFAR100 & 70.7 & 91.3\\
		Vgg19bn & TinyImageNet & 60.1 & 80.2\\
		\bottomrule
	\end{tabular*}
\end{table}
\par
We evaluate the effectiveness of our method using ChangeRate as mentioned in the formula (\ref{eq5}). And we set different $\epsilon$ when using different data iterations into fragile samples. $\epsilon$ is 1e-4 in MNIST and TinyImageNet, and 1e-5 in Cifar.\par
Since backdoor attacks and data poisoning are required to fine-tune the model for attack purposes by using the contaminated data. We fine-tune it in our experiments using normal training samples as well as lower learning rates to detect such tuning compared to it would be more stringent.We also set up a model weight modification experiment by randomly assigning small amplitudes of Gaussian noise to the model weights. The magnitude of these added noises is so small that it does not even affect the performance of the model on the test set.\par
We run our expirements on a host machine with Nvidia 2070 GPU, AMD Ryzen 5 3600X and 16GB memory. Under this setting, it takes about 160 seconds to iterate through a TinyImageNet data 10,000 times.
\clearpage
\clearpage
\subsection{Comparison with Related work}
\begin{table*}[htp]
	\begin{center}
		\begin{minipage}{\textwidth}
			\caption{The table compares our method with previous fragile model watermarking methods, highlighting that our method is a decision-based black-box approach during watermarking and a fully black-box method during validation, with no performance impact on the target model. 
				In comparison to ICIP, another decision-based black-box method, the experiments in Section \ref{sec4.3} have demonstrated that our fragile samples exhibit far greater sensitivity to model modifications.}\label{Table:11}
			\begin{tabular*}{\textwidth}{@{\extracolsep{\fill}}lccc@{\extracolsep{\fill}}}
				\toprule%
				Method & \makecell{Watermarking\\Method} & \makecell{Verification\\Method} & \makecell{Model\\Performance}  \\
				\midrule
				ACMMM-\cite{fragile1}  & White-box & White-box & $\downarrow$ \\
				INS-\cite{fragile2} & White-box & White-box & $\downarrow$ \\
				KSEM-\cite{fragile3} & White-box & {\bf{Black-box}} & $\downarrow$ \\
				AAAI-\cite{fingerprint4} & White-box & {\bf{Black-box}} & $\downarrow$ \\
				CVPR-\cite{fingerprint1} & White-box & {\bf{Black-box}}  & {\bf{=}}\\
				ACSAC-\cite{fingerprint2} & White-box & {\bf{Black-box}} & {\bf{=}}\\
				DAC-\cite{fingerprint3} & White-box & {\bf{Black-box}} & {\bf{=}}\\
				KNOSYS-\cite{fingerprint5} & White-box & {\bf{Black-box}} & {\bf{=}} \\
				ICIP-\cite{fingerprint6} & {\bf{Decision-based Black-box}} & {\bf{Black-box}} & {\bf{=}} \\
				Ours & {\bf{Decision-based Black-box}} & {\bf{Black-box}} & {\bf{=}} \\
				\bottomrule
			\end{tabular*}
		\end{minipage}
	\end{center}
\end{table*}
\begin{figure*}[htbp]
	\centering
	\includegraphics[width=\textwidth]{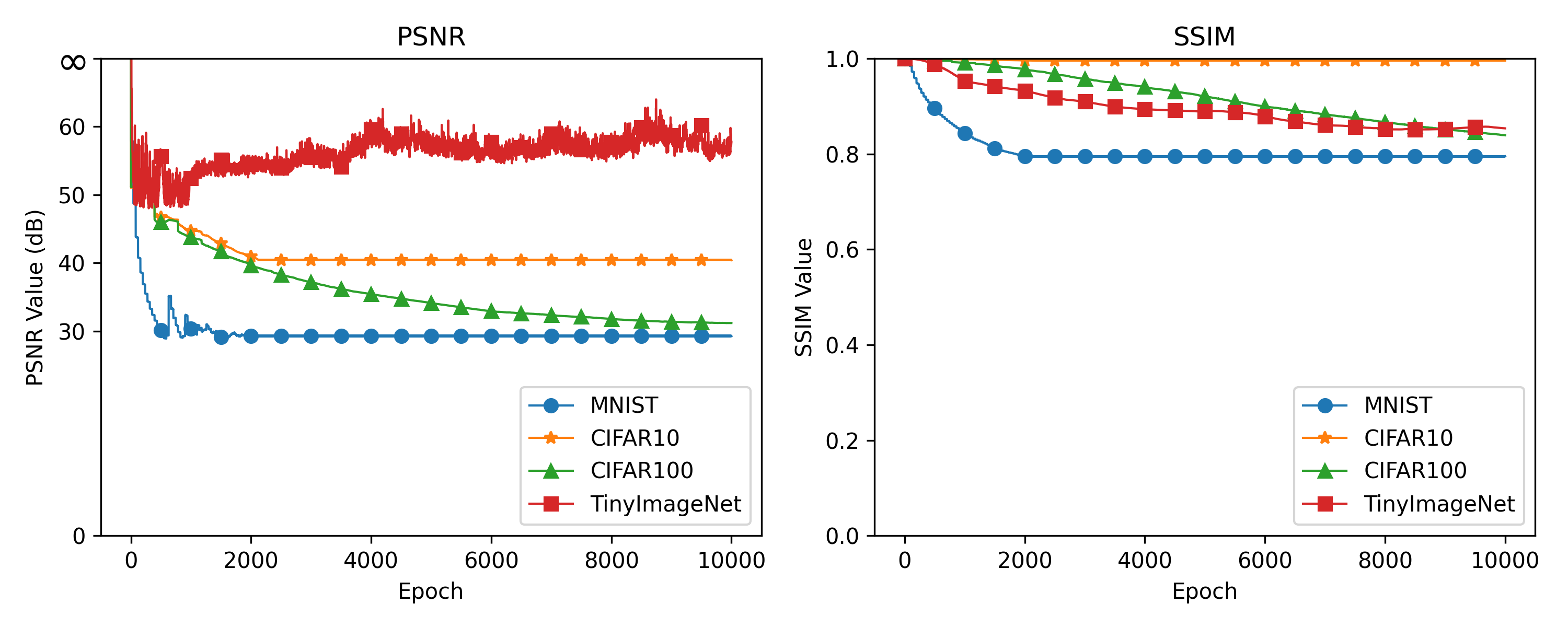}
	\caption{This figure shows the PSNR and structural similarity (SSIM) curves for each dataset during the iteration of the samples into fragile samples. }
	\label{FIG:8}
\end{figure*}
Table \ref{Table:11} shows a comparison of our approach with previous fragile model watermarking approaches. Our approach is decision-based black-box during watermarking, which greatly reduces the adversarial attack risks caused by model information exposure due to frequent iterative updates of the model leading to the need for repeated embedding of watermarks. The approach is completely black-box during the validation period, and only a small number of samples need to be input to complete the validation. And there is no performance impact on the target model throughout the watermark embedding and validation process. Compared with ICIP, which is also a decision-based black-box approach, our fragile samples are not only much better in terms of visual quality, but also much more sensitive to model modifications, as demonstrated in the Section \ref{sec4.3}.\par
\subsection{Image Quality for Fragile Samples}
\begin{figure*}[htbp]
	\centering
	\includegraphics[width=\textwidth]{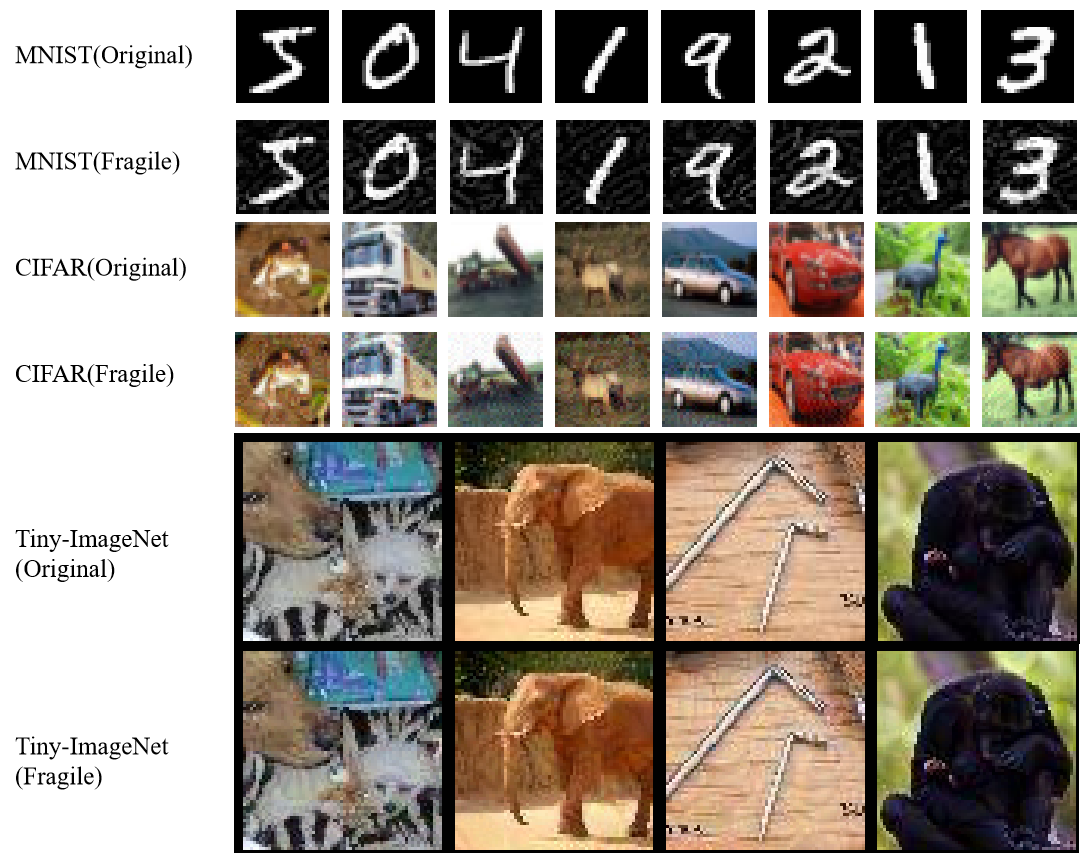}
	\caption{This figure illustrates the visual comparison between the original and fragile images in different datasets. Fragile images generated from MNIST have a lower PSNR of 25 dB, while those generated from more complex color images like CIFAR and TinyImageNet achieve higher PSNRs of 39 dB and 55 dB, respectively.
	The one-dimensional MNIST fragile image can clearly see the perturbations compared to the original image, the three-dimensional CIFAR fragile image can only see some less obvious noise-like perturbations compared to the original image, and the TinyImageNet fragile image, which is twice the size of CIFAR, can barely perceive the difference when compared to the original image.}
	\label{FIG:5}
\end{figure*}
In this section, we focus on the visual quality of the fragile samples and the iterative process of the samples. 
Figure \ref{FIG:8} shows the curves of PSNR and structural similarity\citep{ssim} (SSIM) during the iteration of each dataset sample into fragile samples. The values of PSNR and SSIM for the MNIST data eventually stabilize around 29 dB and 0.8, respectively. Also for the Cifar data, the SSIM of the Cifar10 data ended up being much higher than that of Cifar100, and the PSNR of Cifar10 was able to reach around 40 dB, while the PSNR of Cifar100 ended up being only around 30 dB. We believe that this is because Cifar10 has much fewer categories compared to CIfar100, so the Cifar100 data would be a bit more perturbed to iterate into fragile samples with uniform prediction probabilities for each category.
Finally, the TinyImageNet data, whose loss is close to convergence, can have a PSNR as high as around 60 dB, also because its PSNR value is at a higher level, it is also more prone to oscillation in the next round of perturbation.
Figure \ref{FIG:5} shows the visual comparison between the original image and the fragile image in these datasets.
We believe that larger size images contain more information and have larger capacity, so the increased perturbation will appear more subtle in each iteration.
\par
\begin{figure*}[htbp]
	\centering
	\includegraphics[width=\textwidth]{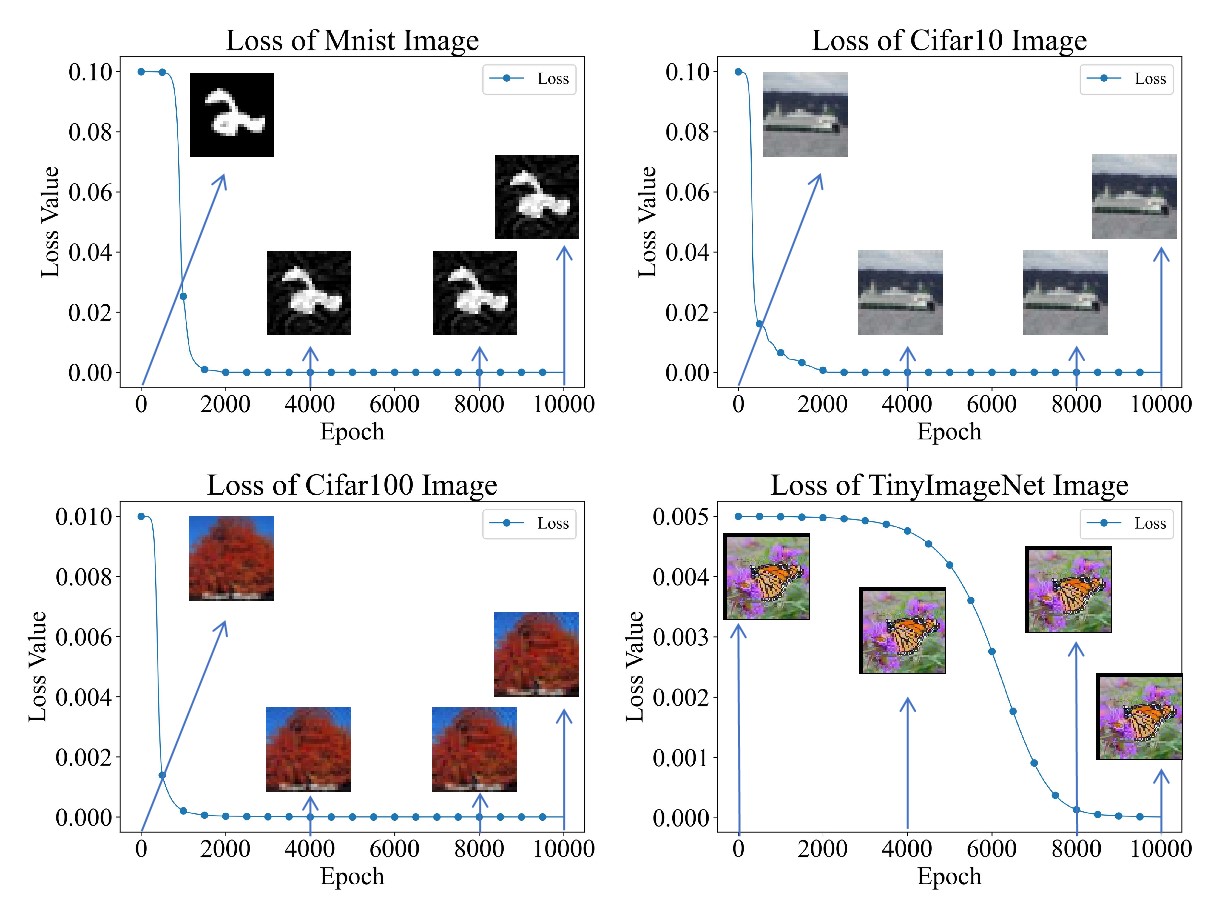}
	\caption{The figure shows how the loss of our algorithm changes as it gradually iterates normal samples into fragile samples. For consistency, one sample randomly selected from each dataset is iterated for 10,000 epochs. We intercepted the images of the samples at epochs 0, 4000, 8000, and 10000 and show them. It can be seen that samples of smaller size require fewer rounds to converge compared to samples of larger size.}
	\label{FIG:6}
\end{figure*}
\begin{figure*}[htbp]
	\centering
	\includegraphics[width=\textwidth]{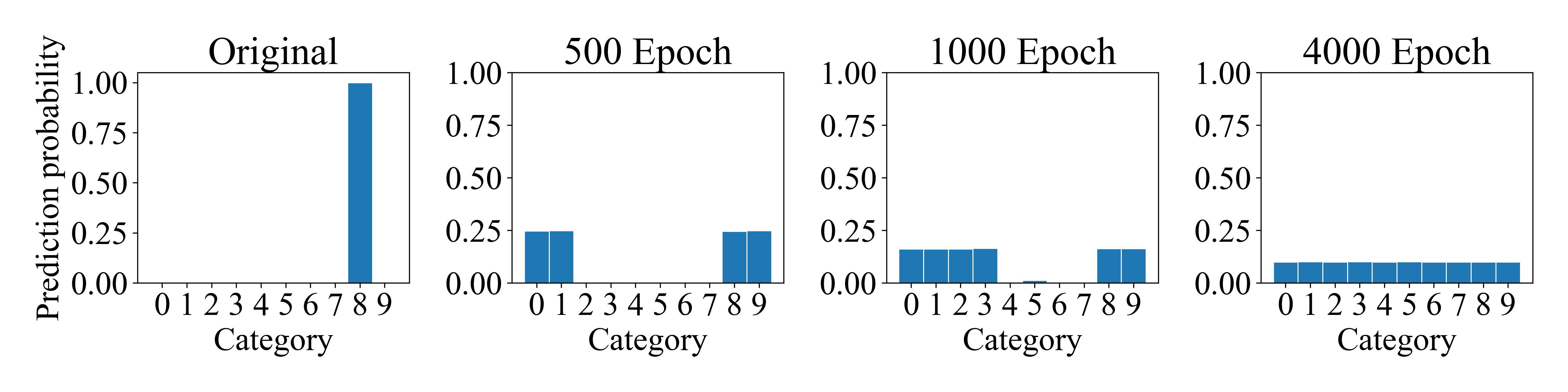}
	\caption{This figure illustrates the process of the probability distribution change of a normal sample into a fragile sample in the target model when our algorithm iterates that fragile sample. To demonstrate this process, we randomly selected a sample from CIFAR10. As shown in the original image on the far left, the target model initially predicts with high confidence that the sample belongs to category 8. After multiple rounds of iterations, the target model becomes increasingly uncertain and imprecise in its predictions, being fuzzy on categories 0, 1, 8, and 9 at round 500, and on more categories at round 1000, until it finally converges and becomes fuzzy on all categories.}
	\label{FIG:7}
\end{figure*}
Subsequently, we randomly selected a sample from each dataset. 
The process of iterating these samples into fragile samples step by step is recorded in detail. 
The process of their loss variation is shown in Figure \ref{FIG:6}. 
For consistency, we set the number of iterations to all 10,000 epochs, even though the loss has long converged by the 4000th epoch for simple pictures like MNIST. 
We intercepted the images of the samples at epochs 0, 4000, 8000, and 10000 and show them. 
It can be seen that samples of smaller size require fewer rounds to converge compared to samples of larger size.
\par
Figure \ref{FIG:7} illustrates the process of the probability distribution change of a normal sample into a fragile sample in the target model when our algorithm iterates that fragile sample. To demonstrate this process, we randomly selected a sample from CIFAR10. As shown in the original image on the far left, the target model initially predicts with high confidence that the sample belongs to category 8. After multiple rounds of iterations, the target model becomes increasingly uncertain and imprecise in its predictions, being fuzzy on categories 0, 1, 8, and 9 at round 500, and on more categories at round 1000, until it finally converges and becomes fuzzy on all categories.
\clearpage
\clearpage
\subsection{Integrity Attack via Weight Modification}\label{sec4.3}
\begin{figure*}[htbp]
	\centering
	\includegraphics[width=\textwidth]{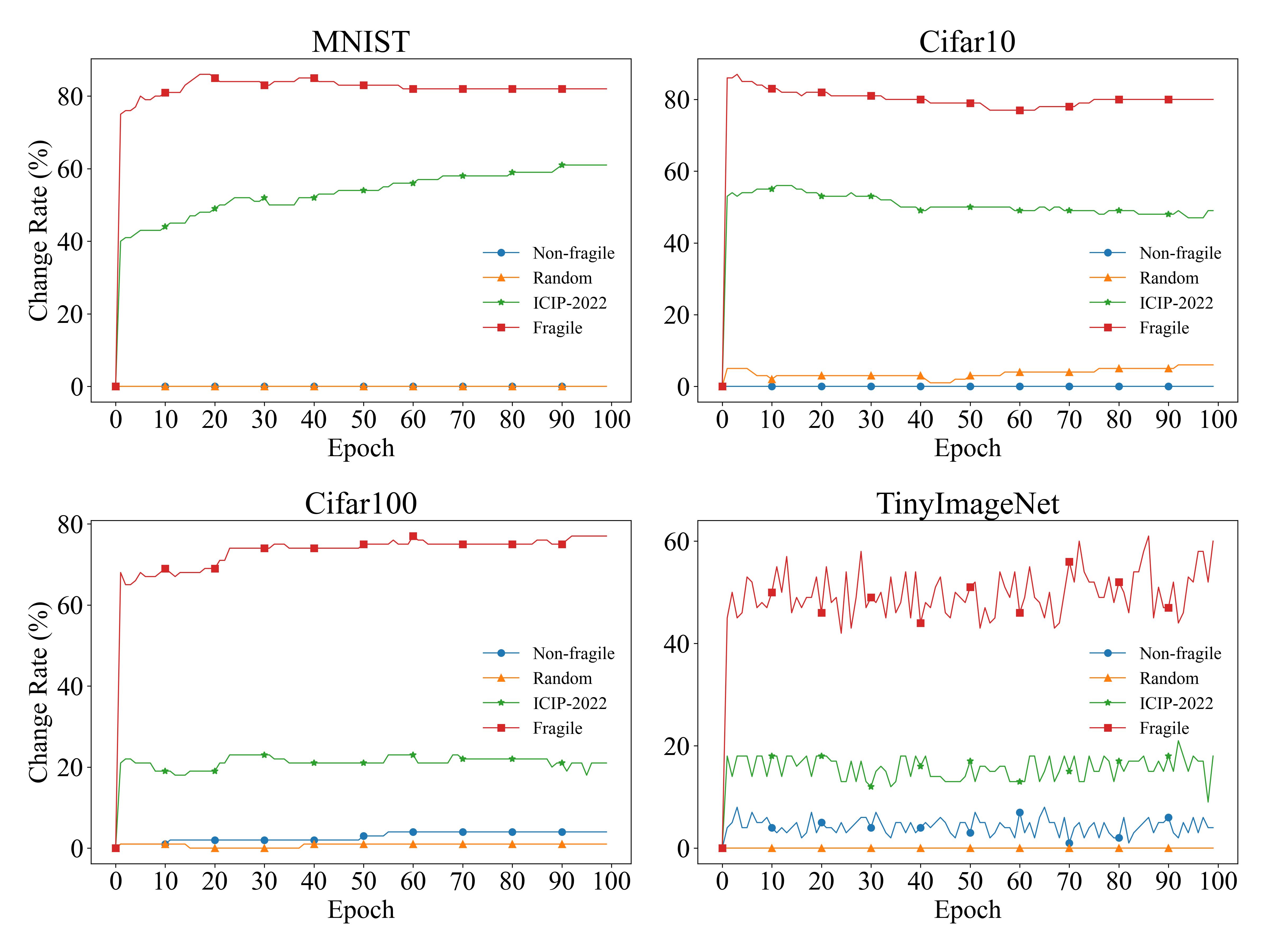}
	\caption{The figure shows the ChangeRate curves for different baseline samples and fragile samples obtained using our method when subjected to integrity attacks such as model fine-tuning. It is clear that the ChangeRate of normal training samples and randomly generated noisy samples do not show significant changes with fine-tuning. However, the fragile samples generated by our method exhibit higher ChangeRate during model fine-tuning compared to the ICIP-2022 method.}
	\label{FIG:9}
\end{figure*}
\begin{figure}[htbp]
	\centering
	\includegraphics[width=0.45\textwidth]{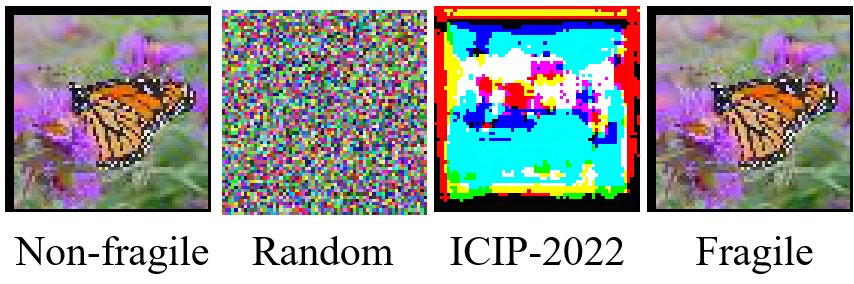}
	\caption{This figure shows the image from non-fragile samples, random noise, the fragile sample generated by the ICIP-2022 Method and our fragile sample.}
	\label{FIG:10}
\end{figure}

In this section, we focus on evaluating the detection efficiency of our method in the face of various integrity attack scenarios. 
Non-fragile samples, randomly generated noise samples, and ICIP-2022 \citep{fingerprint6} methods are selected as baselines for comparison. 
Non-fragile samples were randomly selected from the test set.
Figure \ref{FIG:10} shows these four samples, and the ICIP-2022 method generates fragile samples that the human eye is unable to identify exactly what they are.\par
In calculating the rate of change of sample prediction results for each sample for model modification, we selected 100 samples from each of the samples obtained by each baseline method and selected SGD as the optimizer and the learning rate was set to 1e-3 to fine-tune the model for a total of 100 rounds. The experimental results are shown in Figure \ref{FIG:9}. The ChangeRate of normal training samples and randomly generated noisy samples is extremely low, almost close to 0\%. The fragile samples obtained by the algorithm in this paper have a much higher ChangeRate than the fragile samples generated by the ICIP-2022 method.\par
Thus our method obtains samples that are not only visually much better than the ICIP-2022 method, but also obtains fragile samples that are much more sensitive to model changes than the ICIP-2022 method.
\par
We next demonstrate the effectiveness of our method for testing model integrity using a more stringent modification.
\clearpage
\clearpage
\begin{table}[width=.45\textwidth,cols=4,pos=ht]
	\caption{On the MNIST dataset, we fine-tuned the model using different learning rates and calculated the average ChangeRate for each method.}
	\label{Table:3}
	\begin{tabular*}{\tblwidth}{@{} LLLL@{} }
		\toprule
		Learning Rate & 0.001 & 0.0001 & 0.00001 \\
		\midrule
		Non-fragile & 0.06 & 0.06 & 0.06 \\
		Random & 0.00 & 0.00 & 0.00 \\
		ICIP-2022 & 45.00 & 40.00 & 38.00 \\
		Ours & {\bf{80.00}} & {\bf{75.00}} & {\bf{74.00}} \\
		\bottomrule
	\end{tabular*}
\end{table}
\begin{table}[width=.45\textwidth,cols=4,pos=ht]
	\caption{On the Cifar10 dataset, we fine-tuned the model using different learning rates and calculated the average ChangeRate for each method.}
	\label{Table:4}
	\begin{tabular*}{\tblwidth}{@{} LLLL@{} }
		\toprule
		Learning Rate & 0.001 & 0.0001 & 0.00001 \\
		\midrule
		Non-fragile & 2.05 & 1.40 & 1.21 \\
		Random & 15.00 & 5.00 & 3.00 \\
		ICIP-2022 & 48.00 & 30.00 & 21.00 \\
		Ours & {\bf{80.00}} & {\bf{77.00}} & {\bf{75.00}} \\
		\bottomrule
	\end{tabular*}
\end{table}
\begin{table}[width=.45\textwidth,cols=4,pos=ht]
	\caption{On the Cifar100 dataset, we fine-tuned the model using different learning rates and calculated the average ChangeRate for each method. }
	\label{Table:5}
	\begin{tabular*}{\tblwidth}{@{} LLLL@{} }
		\toprule
		Learning Rate & 0.001 & 0.0001 & 0.00001 \\
		\midrule
		Non-fragile & 6.13 & 5.61 & 4.91 \\
		Random & 3.00 & 3.00 & 3.00 \\
		ICIP-2022 & 22.00 & 21.00 & 19.00 \\
		Ours & {\bf{89.00}} & {\bf{86.00}} & {\bf{83.00}} \\
		\bottomrule
	\end{tabular*}
\end{table}
\begin{table}[width=.45\textwidth,cols=4,pos=ht]
	\caption{On the TinyImageNet dataset, we fine-tuned the model using different learning rates and calculated the average ChangeRate for each method. }
	\label{Table:6}
	\begin{tabular*}{\tblwidth}{@{} LLLL@{} }
		\toprule
		Learning Rate & 0.001 & 0.0001 & 0.00001 \\
		\midrule
		Non-fragile & 3.83 & 3.10 & 2.90 \\
		Random & 0.00 & 0.00 & 0.00 \\
		ICIP-2022 & 13.00 & 13.00 & 12.00 \\
		Ours & {\bf{53.00}} & {\bf{51.00}} & {\bf{50.00}} \\
		\bottomrule
	\end{tabular*}
\end{table}
To test the sensitivity of our fragile samples to smaller modifications in model fine-tuning, we fine-tuned the model using lower learning rates on four different combinations of datasets. We used all 10,000 test set data as non-fragile data to more accurately represent the ChangeRate of such non-fragile samples after each round of fine-tuning. Our results, as presented in Tables \ref{Table:3}, \ref{Table:4}, \ref{Table:5}, and \ref{Table:6}, clearly show that the fragile samples obtained through our method are more sensitive to model fine-tuning.
\par
\begin{table}[width=.45\textwidth,cols=4,pos=ht]
	\caption{On the MNIST dataset, we add a Gaussian noise with different standard deviation on overall model parameters and calculated the average ChangeRate for each method.}
	\label{Table:7}
	\begin{tabular*}{\tblwidth}{@{} LLLL@{} }
		\toprule
		Standard Deviation & 0.001 & 0.0005 & 0.0001 \\
		\midrule
		Non-fragile & 0.03 & 0.03 & 0.00 \\
		Random & 0.00 & 0.00 & 0.00 \\
		ICIP-2022 & 6.00 & 1.00 & 0.00 \\
		Ours & {\bf{86.00}} & {\bf{84.00}} & {\bf{82.00}} \\ 
		\bottomrule
	\end{tabular*}
\end{table}
\begin{table}[width=.45\textwidth,cols=4,pos=ht]
	\caption{On the Cifar10 dataset, we add a Gaussian noise with different standard deviation on overall model parameters and calculated the average ChangeRate for each method.}
	\label{Table:8}
	\begin{tabular*}{\tblwidth}{@{} LLLL@{} }
		\toprule
		Standard Deviation & 0.001 & 0.0005 & 0.0001 \\
		\midrule
		Non-fragile & 0.74 & 0.39 & 0.05 \\
		Random & 9.00 & 0.00 & 0.00 \\
		ICIP-2022 & 17.00 & 3.00 & 2.00 \\
		Ours & {\bf{83.00}} & {\bf{75.00}} & {\bf{70.00}} \\
		\bottomrule
	\end{tabular*}
\end{table}
\begin{table}[width=.45\textwidth,cols=4,pos=ht]
	\caption{On the Cifar100 dataset, we add a Gaussian noise with different standard deviation on overall model parameters and calculated the average ChangeRate for each method.}
	\label{Table:9}
	\begin{tabular*}{\tblwidth}{@{} LLLL@{} }
		\toprule
		Standard Deviation & 0.005 & 0.0025 & 0.001 \\
		\midrule
		Non-fragile & 9.93 & 5.66 & 2.12 \\
		Random & 12.00 & 4.00 & 4.00 \\
		ICIP-2022 & 62.00 & 6.00 & 1.00 \\
		Ours & {\bf{86.00}} & {\bf{82.00}} & {\bf{56.00}} \\
		\bottomrule
	\end{tabular*}
\end{table}
\begin{table}[width=.45\textwidth,cols=4,pos=ht]
	\caption{On the TinyImageNet, we add a Gaussian noise with different standard deviation on overall model parameters and calculated the average ChangeRate for each method.}
	\label{Table:10}
	\begin{tabular*}{\tblwidth}{@{} LLLL@{} }
		\toprule
		Standard Deviation & 0.001 & 0.0005 & 0.0001 \\
		\midrule
		Non-fragile & 13.49 & 6.90 & 1.47 \\
		Random & 0.00 & 0.00 & 0.00 \\
		ICIP-2022 & 82.00 & 31.00 & 5.00 \\
		Ours & {\bf{94.00}} & {\bf{76.00}} & {\bf{22.00}} \\
		\bottomrule
	\end{tabular*}
\end{table}
To evaluate the sensitivity of our fragile samples to small modifications, we followed the convention of previous method tests and added random noise with mean 0 and different standard deviation to the parameters of the model. As presented in Tables \ref{Table:7}, \ref{Table:8}, \ref{Table:9}, and \ref{Table:10}, we set the standard deviation to be extremely small, making it nearly impossible for an attacker to achieve an attack effect by modifying the model to such a degree. Nonetheless, our method still exhibited high sensitivity to these small modifications, even when other methods such as ICIP-2022 failed. This further validates the efficacy of our approach.
\clearpage

\section{Conclusion}\label{}
In this paper, we propose a decision-based iterative fragile watermarking algorithm, which converts normal training samples into fragile samples that are sensitive to model changes, enabling the detection of model compromise. 
The proposed algorithm is an optimization problem that minimizes the variance of the predicted probability distribution of the target model. 
Fragile samples are generated by iterating over normal samples, and our method offers several advantages: 
(1) the iterative update of samples is done in a decision-based black-box manner, relying solely on the predicted probability distribution of the target model, which reduces the risk of exposure to adversarial attacks; 
(2) the small-amplitude multiple iterations approach allows the fragile samples to exhibit superior visual quality, with a PSNR of 55 dB in TinyImageNet compared to the original samples; 
(3) even with changes in the overall parameters of the model of magnitude 1e-4, the fragile samples can detect such changes; 
and (4) the method is independent of the specific model structure and dataset. We demonstrate the effectiveness of our method on multiple models and datasets, and show that it outperforms the current state-of-the-art methods.
\par
But as AI is moving towards larger models that can handle multimodal tasks, the focus of future research should shift accordingly. Specifically, there is a need to address potential security risks associated with large multimodal models and contribute to enhancing the overall security environment of AI.
\section*{Acknowledgement}
This work is supported by National Natural Science Foundation of China
under Grant No.62172001, U1936214.
\section*{CRediT authorship contribution statement}
{\bf{Heng Yin:}} Writing, original draft, Editing, Revision and modification, Investigation. {\bf{Zhaoxia Yin:}}Revision and modification, Supervision, Reviewing and editing. {\bf{Zhenzhe Gao:}}Revision and modification, Reviewing and editing. {\bf{Hang Su:}}Revision and modification, Supervision, Reviewing and editing. {\bf{Xinpeng Zhang:}}Revision and modification, Supervision, Reviewing and editing.
\section*{Declaration of competing interest}
The authors declare that they have no known competing financial interests or personal relationships that could have appeared to influence the work reported in this paper.


\end{document}